\documentclass[pra,amsmath,twocolumn,superscriptaddress,showpacs]{revtex4-1}

\usepackage{epsfig}

\usepackage{amssymb}
\usepackage{amsmath}
\usepackage{amsfonts}
\usepackage{bbold}
\usepackage{bm}
\usepackage{graphicx}
\usepackage{braket}
\usepackage{color}
\usepackage{hyperref}

\begin{document}
\title{Measuring the Edwards-Anderson order parameter of the Bose glass: \\ a quantum gas microscope approach}
\author{S.~J. Thomson}
\altaffiliation[Present Address: ]{Institut de Physique Th\'eorique, CEA Saclay, 91191 Gif-sur-Yvette, France}
\email{steven.thomson@cea.fr}
\affiliation{SUPA, School of Physics and Astronomy, University of St.~Andrews, North Haugh, St. Andrews, KY16 9SS, United Kingdom}
\author{L.~S. Walker}
\altaffiliation[Present Address: ]{University of Strathclyde, Department of Physics, SUPA, Glasgow G4 0NG, United Kingdom}
\affiliation{SUPA, School of Physics and Astronomy, University of St.~Andrews, North Haugh, St. Andrews, KY16 9SS, United Kingdom}
\author{T.~L. Harte}
\affiliation{Clarendon Laboratory, University of Oxford, Parks Road, Oxford OX1 3PU, United Kingdom}
\author{G.~D. Bruce}
\email{gdb2@st-andrews.ac.uk}
\affiliation{SUPA, School of Physics and Astronomy, University of St.~Andrews, North Haugh, St. Andrews, KY16 9SS, United Kingdom}

\date{\today}

\begin{abstract}
With the advent of spatially resolved fluorescence imaging in quantum gas microscopes, it is now possible to directly image glassy phases and probe the local effects of disorder in a highly controllable setup. Here we present numerical calculations using a spatially-resolved local mean-field theory, show that it captures the essential physics of the disordered system and use it to simulate the density distributions seen in single-shot fluorescence microscopy. From these simulated images we extract local properties of the phases which are measurable by a quantum gas microscope and show that unambiguous detection of the Bose glass is possible. In particular, we show that experimental determination of the Edwards-Anderson order parameter is possible in a strongly correlated quantum system using existing experiments. We also suggest modifications to the experiments which will allow further properties of the Bose glass to be measured.
\end{abstract}


\maketitle

\section{Introduction}

Ultracold atoms in optical lattices provide a highly controllable environment for quantum simulations of real materials \cite{Bloch+12} and in particular for isolating and investigating the effects of disorder. The role of disorder has already been investigated in cold atom systems exhibiting Anderson localization \cite{Billy+08,Roati+08,Kondov+11,Jendrzejewski+12,Semeghini+15}, many-body localization \cite{Schreiber+15,Bordia+16,Choi+16} and quantum glass phases \cite{Fallani+07,White+09,Deissler+10,Pasienski+10,Gadway+11,Deissler+12,Tanzl+13,DErrico+14,Meldgin+16}. Most experiments to date have been measurements of bulk properties, such as coherence and transport, however the single-site resolved detection of bosonic \cite{Bakr+09,Sherson+10} and fermionic \cite{Haller+15,Cheuk+15,Parsons+15,Omran+15,Edge+15} atoms now affords us the ability to investigate the local properties of strongly correlated systems \cite{Kuhr+16}. Very recently, quantum gas microscopes have been used to investigate transport in disordered systems in the context of many-body localisation \cite{Choi+16}, with measurements of entanglement entropy on the horizon \cite{Islam+15,Kaufman+16}.

No such local measurements have yet been performed to study the Bose glass phase \cite{Giamarchi+88,Fisher+89} or any of its fermionic analogues despite quantum gas microscopes offering a natural environment in which to investigate the local effects of disorder so important to the physics of these phases. Here, we seek to motivate the site-resolved investigation of glassy phases and show that quantum gas microscopes are ideally suited to measuring local quantities of disordered systems. In particular, we show that quantum gas microscopes are capable of measuring the Edwards-Anderson order parameter.

This quantity has never been measured experimentally in any condensed matter system as it requires detailed knowledge of the microscopic states of individual lattice sites. Before the development of quantum gas microscopes, Morrison \emph{et al.} \cite{Morrison+08} proposed an as-yet-unrealized method to extract the Edwards-Anderson order parameter of the Bose glass by generating two independent copies of the system with the same disorder distribution, physically overlapping the two copies and measuring the bulk properties. Here, we suggest an alternative quantum gas microscope approach to measuring it.

Using the disordered Bose-Hubbard model in two dimensions we simulate ultracold bosons in an optical lattice and perform mean-field calculations of the lattice occupation across a range of tunneling and chemical potential values. We map out the phase diagram in terms of the Edwards-Anderson order parameter to show that it is capable of distinguishing the Bose glass and demonstrate that it can be measured under realistic experimental conditions of parity-sensitive detection, harmonic confinement and finite temperature. 

\section{Model}

The Bose-Hubbard model describes spinless bosons on a hypercubic lattice and has been shown to be a good description of ultracold atoms in optical lattices \cite{Jaksch+98}. In the presence of chemical potential disorder the Hamiltonian is given by
\begin{equation}
\mathcal{H} = -J \sum_{\langle i, j \rangle} \left( \hat{b}^{\dagger}_{i} \hat{b}_{j} + \hat{b}^{\dagger}_{j} \hat{b}_{i}\right) + \sum_{i} \left[ \frac{U}{2} \hat{n}_{i} (\hat{n}_{i}-1) -\mu_i \hat{n}_{i} \right],
\end{equation}
where $\hat{n}_i$, $\hat{b}^{\dagger}_i$ and $\hat{b}_{i}$ respectively count, create and destroy particles on site $i$. $J$ is the tunneling amplitude between nearest-neighbour sites, $U$ is the on-site interaction and $\mu_i=\mu+\varepsilon_i$ where $\mu$ is the bulk chemical potential \footnote{By treating the central region of the harmonic trap as the ``core'' region and the region around the edges of the trap as a particle reservoir similarly to Ref. \cite{Roscilde+10}, we can use calculations in the grand canonical ensemble to describe the region of interest at the trap center.} and $\varepsilon_i$ is a spatially uncorrelated random variable drawn from a symmetric box distribution of width $2 \Delta$ which describes the disorder.

Disorder of this type can be approximated either by superimposing a speckle potential on top of the lattice \cite{Billy+08,White+09} or by using a spatial light modulator \cite{Bruce+15,Choi+16} to vary the lattice depth from site-to-site. In a real experiment, adding any form of disorder will simultaneously
modify not only the on-site chemical potential, but also the hopping amplitude and the strength of the on-site repulsion. In the following, we assume that the dominant effects can be modeled by local chemical potential disorder alone.

In the clean case ($\varepsilon_i=0$), the model contains two phases. In the non-interacting limit ($U \rightarrow 0$) the ground state is a gapless, compressible superfluid (SF), while in the local limit ($J \rightarrow 0$) the ground state is a Mott insulator (MI). In presence of disorder ($\varepsilon_i \neq 0$), a Bose glass (BG) phase always intervenes between the MI and the SF \cite{Pollet+09,Gurarie+09}. The BG is a Griffiths phase \cite{Vojta10} characterized by the presence of rare disconnected superfluid regions within a Mott-insulating background and is a gapless, compressible insulating phase. It has been the subject of extensive theoretical work using techniques ranging from mean-field theories \cite{Buonsante+07,Bissbort+09,Bissbort+10,Stasinska+12} to renormalisation group \cite{Giamarchi+88,Fisher+89,Mukhopadhyay+96,Herbut97,Herbut98,Weichman+08,Kruger+09,Kruger+11,Hegg+13,Thomson+14} to quantum Monte Carlo and other numerical methods \cite{Scalettar+91,Krauth+91,Kisker+97,Sen+01,Lee+01,Prokof'ev+04,Niederle+13}. Though experimental work on the Bose glass has so far concentrated on bulk properties, many theoretical works consider local properties which have not yet been experimentally measured. 

In order to motivate the investigation of the local properties of disordered systems using quantum gas microscopes, here we simulate measurements of the Edwards-Anderson order parameter, a local property accessible to current quantum gas microscope experiments which can be straightforwardly computed across a wide range of parameter values for comparison with experiments. 

\section{Edwards-Anderson Order Parameter}

By analogy with spin glass systems \cite{Binder+86,Fischer+91}, various Edwards-Anderson-like order parameters for the Bose glass have been proposed \cite{Morrison+08,Graham+09,Thomson+14,Khellil+16}. The Edwards-Anderson order parameter originally arose in the mean-field theory of spin glasses as an indicator of the non-trivial breaking of ergodicity in a disordered system. It is the natural order parameter for a disordered phase, and is a more appropriate metric for distinguishing the Bose glass than, for example, compressibility, which can be induced by both disorder and temperature.

Here we define the Edwards-Anderson order parameter in terms of the boson number density as
\begin{equation}
q=\overline{\langle \hat{n}_{i} \rangle^{2}}-\overline{\langle \hat{n}_i \rangle}^{2},
\label{eqn:q}
\end{equation}
where the angled brackets refer to the thermal average and the overline refers to the disorder average.

By construction, this disorder-averaged correlation function is identically zero everywhere in the clean system. The Mott insulator is characterized by an integer value of $\langle \hat{n}_i \rangle$ on every site and consequently a vanishing $q$. The homogeneous superfluid is characterized by a uniform but non-integer $\langle \hat{n}_i \rangle$ which also leads to $q=0$.

In the disordered system, however, any correlation between the density and the disorder will lead to a non-zero value of $q$. This in principle allows the Edwards-Anderson order parameter to distinguish between the homogeneous MI and SF phases where $q=0$ and the BG phase where $q \neq 0$. However, in the presence of the chemical potential disorder most conveniently realized in experiments, the superfluid phase that emerges from the BG will be inhomogeneous. Consequently, it will also exhibit a non-zero value of $q$, albeit a smaller value than in the BG due to the reduced correlation between disorder and density. This limits the usefulness of the order parameter in distinguishing, for example, the BG-SF transition. However an appropriately chosen disorder distribution (such as a bimodal distribution or pure hopping disorder) should alleviate or even eliminate this problem. Nonetheless, in the following we restrict ourselves to random chemical potential disorder as this is the type most easily added to current experiments.

\section{Local mean-field Theory}

We employ a spatially resolved local mean-field theory to simulate the experimental results, using the Gutzwiller variational wavefunction
\begin{equation}
\ket{\Psi} = \prod_{i} \sum_{n_i} \frac{f_{n_{i},i}}{\sqrt{n_i!}}(\hat{b}^{\dagger}_{i})^{n_i} \ket{0},
\end{equation}
subject to the normalisation constraint $\sum_{n_i} |f_{n_i,i}|^{2}=1 \medspace \forall i$. This provides us with a variational energy in terms of the parameters $f_{n_i,i}$ given by $E_{MF}=\braket{\Psi|\mathcal{H}|\Psi}$ which we minimize using a conjugate gradient algorithm \cite{Harte+14}.
This wavefunction has been shown to provide a good qualitative description of interacting bosons in both clean \cite{Rokhsar+91,Jaksch+98} and disordered systems \cite{Buonsante+07,Stasinska+12,Yan+16}. Here we simulate experiments in two dimensions where local mean-field theory is a reliable and accurate method. While techniques exist which are able to extract bulk properties in the thermodynamic limit without recourse to the calculation of local properties, such as stochastic mean-field theory \cite{Bissbort+09,Bissbort+10}, here it is precisely the local behavior of small systems which we are interested in replicating in order to meaningfully compare with experiments.

In the clean case, the minimization procedure always finds the global minimum. In the presence of disorder, the energy landscape can become complicated due to the presence of multiple local minima so it is necessary to check convergence by testing multiple different initial configurations. For the regions of the phase diagram which are experimentally accessible, we find that it is sufficient to truncate the ansatz wavefunction at $n_{i}=6$. After obtaining the values of the variational coefficients $f_{n_i,i}$ for each site, we use them to probabilistically calculate the parity-limited occupancy on each site and generate simulated ``snapshot'' density distributions that mimic those produced in quantum gas microscopes. The probability of imaging site $i$ as empty is $\sum_{n_{i}}^{\text{even}} \left|f_{n_{i},i}\right|^{2}$ while the probability of the site being occupied in the image is $\sum_{n_{i}}^{\text{odd}} \left|f_{n_{i},i}\right|^{2}$. In all of the following, before performing any analysis we first simulate a snapshot image of the lattice as would be seen in a quantum gas microscope, to ensure that the extraction of $q$ can be made in an experimentally realistic number of repetitions.

\section{Mean-Field Phase Diagram} \label{sect:flat}

\begin{figure}[t!]
\begin{center}
\includegraphics[width= \linewidth]{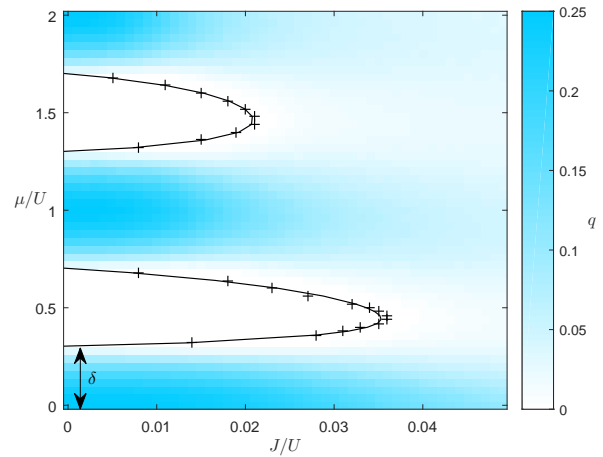}
\caption{(Color online) Phase diagram of the disordered Bose-Hubbard model, showing values of the Edwards-Anderson order parameter $q$, sampled at a resolution of $50 \times 50$ points. Each point was calculated using 10 thermal configurations and 10 disorder realisations on a $25 \times 25$ lattice. The crosses indicate the boundary of the Mott lobes at $q=0$ and the guide-to-the-eye solid line is a fit of these points. The cross size shows the uncertainty due to our sampling resolution. The same color scale is used in Figures \ref{fig:snapshots}, \ref{fig:Clean+Disorder} and \ref{fig:T}. \label{fig:phase}}
\end{center}
\end{figure}

We construct the phase diagram shown in Fig. \ref{fig:phase} for a homogeneous $25\times25$-site lattice superimposed with a disorder distribution of width $\delta=\Delta/U = 0.3$. Comparing the value of $q$ extracted from snapshots of a $25\times 25$ lattice with snapshots from a $100\times 100$ lattice results in an increase in accuracy on the order of a few percent but at the expense of a large increase in computational time. For each point in the phase diagram, we extract 10 snapshots for each of 10 distinct disorder distributions to calculate $q$. We see clear regions of $q \gg 0$ which we identify as the BG surrounding regions of $q = 0$ which we identify as the MI. By contrast, in the case of zero disorder, $q = 0$ across the entire phase diagram.

Strictly, $q$ is only zero in the MI regions because we do not consider the effects of quantum fluctuations. These have been shown \cite{Morrison+08} to lead to a small non-zero $q$ even in the MI, however there remains a sharp crossover from the MI to the BG from which the transition can be determined, confirming that this remains a good order parameter beyond the mean-field analysis presented here.

As we move from the BG to the SF phase, $q$ smoothly goes to zero with no feature indicative of the BG-SF transition. This is because even after the establishment of global superfluid coherence, the chemical potential disorder still causes lingering inhomogeneities in the superfluid. The same result was also found in Ref. \cite{Morrison+08}. To obtain a sharp transition from the BG to the SF using this order parameter, we require a type of disorder which does not cause density inhomogeneities in the SF, such as bimodal hopping disorder \cite{Roscilde+07}. This form of disorder has been proposed for a system of ultracold atoms with cavity-mediated interactions \cite{Gopalakrishnan+11}, but not yet implemented in conjunction with a quantum gas microscope. Here we follow the setup of Ref. \cite{Sherson+10} where chemical potential disorder is more straightforwardly incorporated.

Numerically, $q$ appears to be bounded from above by the value of the variance $\kappa=\sum_{ij} \left[\left\langle n_{i} n_{j} \right\rangle-\left\langle n_{i}\right\rangle\left\langle n_{j}\right\rangle\right]$. Previous studies have shown that the variance is suppressed near the tip of the Mott lobes \cite{Buonsante+07, Bissbort+09}, with some suggesting that it vanishes entirely \cite{Wang+15,Thomson+15}. The suppression of $\kappa$ results in the corresponding suppression of $q$ in the vicinity of the tips of the Mott lobes, where we would in any case expect local mean-field theory to break down.

Due to the parity-sensitivity of the fluorescence imaging technique, the value of $\kappa$ and therefore $q$ saturates at a maximum of $0.25$. We recommend that initial experiments be performed in the region $J/U < 1/100$, where the value of $q$ is largest and there should be no SF present, meaning a non-zero $q$ necessarily corresponds to the BG.

\section{Optical lattices with harmonic confinement}

\begin{figure}[t]
\begin{center}
\includegraphics[width= \linewidth]{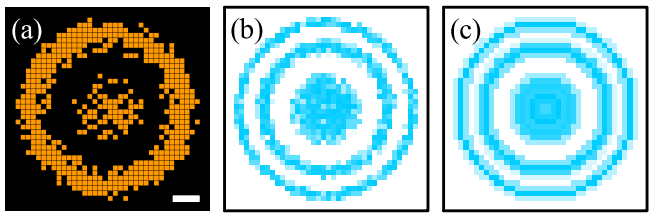}
\caption{(Color online) In-situ measurement of $q$, for $\delta = 0.3$, $\mu_{0}/U=2.1$, and $J/U=1/300$. (a) Simulated snapshot of the parity-sensitive lattice occupation, where orange (black) denotes a site occupied by an odd (even) number of atoms. The white scale bar denotes $5$ lattice sites. (b) $q$ extracted from site-by-site evaluation of Eq. (\ref{eqn:q}) with 10 snapshots at each of 10 disorder realisations. (c) $q$ can also be approximated using 10 repetitions of a single disorder realization by averaging along contours of constant $\mu$. \label{fig:snapshots}}
\end{center}
\end{figure}

In real optical lattice experiments, the Gaussian profile of the laser beams results in a background harmonic confining potential. We emulate the lattice geometry of the experiments detailed in Ref. \cite{Sherson+10}, i.e. we model a $60 \times 60$-site lattice in the low-tunneling regime ($J/U = 1/300$) with trap frequencies $\omega_{x}=\omega_{y}=2\pi \times 77.3~\text{Hz}$ \footnote{This is the geometric mean of $\omega_{x}$ and $\omega_{y}$ used in Ref. \cite{Sherson+10}}. This causes a spatial variation of the chemical potential $\mu\left(r\right) = \mu_{0} −- 0.5 m\left(\omega_{x}^{2} x^{2} + \omega_{y}^{2} y^{2}\right)$, where $m$ is the atomic mass and $\mu_{0}$ the chemical potential in the center of the trap. In a single image of the optical lattice, one captures a range of values of $\mu$ and thus multiple phases. Due to parity-sensitive losses, the familiar ``wedding-cake'' structure of the MI appears as concentric rings of occupied and unoccupied lattice sites. In the presence of disorder, a single image captures both BG and MI regions.

By incorporating $\mu\left(r\right)$ into our Hamiltonian, we generate snapshots of the site occupations in these regions at zero temperature. At $\delta=0$, our model accurately matches the experimental results from Ref. \cite{Sherson+10}. For $\delta=0.3$, as shown in Fig. \ref{fig:snapshots}(a) for $\mu_{0}/U=2.1$, the MI regions are detected as areas with uniform site occupations, while the BG regions are those with non-uniform occupation. There are no SF regions present in the snapshots for these parameter values.

As in Section \ref{sect:flat}, we extract $q$ using Eq. (\ref{eqn:q}) by performing the averages over 10 simulated snapshots at each of 10 disorder realisations [Fig. \ref{fig:snapshots}(b)]. The result gives a clear distinction between the MI regions ($q= 0$) and the BG regions ($q \sim 0.25$) which are present in this low-tunneling regime. The value of $q$ measured using this method of averaging of snapshots agrees to within $1\%$ of that extracted directly from the Gutzwiller coefficients. This method is easily integrated into existing quantum gas microscope experiments such as those in Refs. \cite{Choi+16} and \cite{Zupancic+16}, where the disorder can be generated by a digital micromirror device and thus easily changed to allow averaging over different disorder realizations.

Even without the ability to apply multiple disorder realizations, an indicative measure of $q$ is also possible by using only a single disorder realization, and performing azimuthal averaging along contours of fixed $\mu$ [Fig. \ref{fig:snapshots}(c)]. While this procedure gives qualitatively similar results to the site-averaged case, the regions of non-zero $q$ are broadened and the values differ by $\sim 30\%$ on average when compared to the site-averaged case. These effects are due to the problems inherent in attempting to perform circular averages on a square lattice.

\section{Finite temperature effects}

\begin{figure}[t!]
\begin{center}
\includegraphics[width=\linewidth]{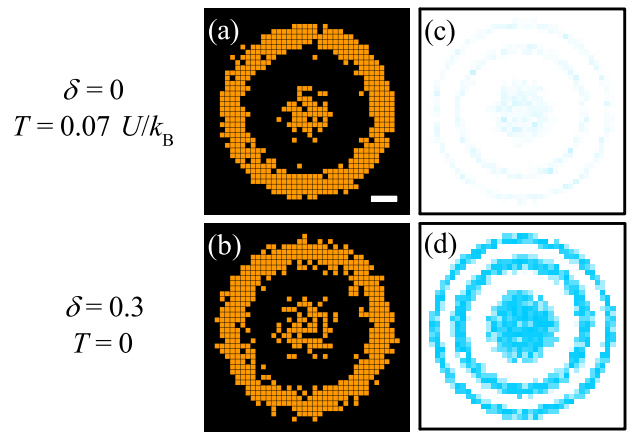}
\caption{\label{fig:Clean+Disorder} (Color online) The site occupations of (a) a clean optical lattice at finite temperature and (b) a disordered lattice at zero temperature are cosmetically similar. However, (c) for the clean case $q\approx0$, while (d) in the disordered case there are Bose glass regions with large values of $q$. The white scale bar shown in (a) denotes $5$ lattice sites. }
\end{center}
\end{figure}

The previous analysis was conducted at zero temperature and did not take into account thermal fluctuations in the density. As density fluctuations due to finite temperature and disorder look cosmetically similar [as shown in Fig. \ref{fig:Clean+Disorder}(a) and (b)], we verify that $q$ can distinguish between these cases. To test this, we model the effect of finite temperature in the limit of zero tunneling by Boltzmann-weighted Fock states. This leads to an additional probabilistic step in the snapshot generation
\begin{equation}
P\left(n\right) = \frac{\exp \left\lbrace\left[\mu\left(r\right)n - E_{n}\right] / k_{B}T\right\rbrace}{Z\left(r\right)},
\end{equation}
\noindent where $T$ is the temperature and $Z\left(r\right)$ the partition function for a homogeneous system with chemical potential $\mu(r)$. Fig. \ref{fig:Clean+Disorder}(c) and (d) show that the site-averaged $q$ gives a clear distinction between thermal fluctuations and glassy phases. The average value of $q$ within a radius of 5 lattice sites from the trap center is $0.02 \pm 0.08$ for the clean, finite temperature case and $0.216 \pm 0.007$ for the disordered, zero-temperature lattice.  The dominant uncertainty stems from the variance in the on-site occupation due to the finite number of simulated snapshots. Consequently, the faint rings of non-zero $q$ visible in Fig. \ref{fig:Clean+Disorder}(c) are an artifact of the finite number of averages performed: by averaging over a large enough number of snapshots, they can be made to disappear entirely. 

\begin{figure}[t!]
\begin{center}
\includegraphics[width= \linewidth]{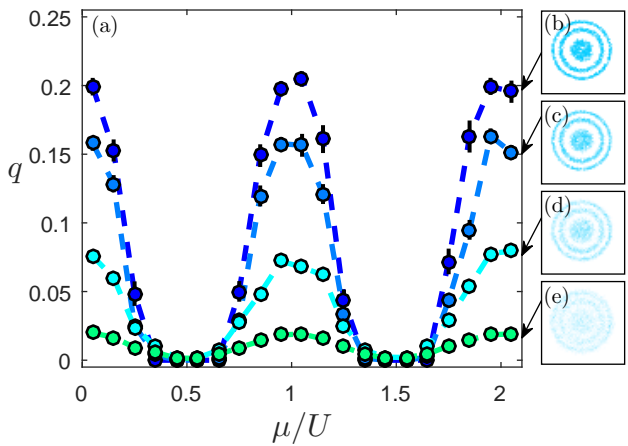}
\caption{(Color online) Melting of the Bose glass. The visibility of the Edwards-Anderson order parameter decreases with temperature but remains finite over a wide range of experimentally-relevant temperatures. (a) Histogram showing the local average value of $q$ with error bars showing standard error. (b) - (d) The full distribution of $q$ in the zero-tunneling regime for $\delta = 0.3$, $\mu_{0}/U=2.1$, and $k_{B}T/U=0.01, 0.05, 0.13$ and $0.25$ respectively.}
\label{fig:T}
\end{center}
\end{figure}

As shown in Fig. \ref{fig:T}, the main effect of increasing temperature at fixed disorder strength is to reduce the maximum value of $q$. Despite this, the Edwards-Anderson order parameter remains distinct across an experimentally relevant range of temperatures. For reference, the lowest temperature reached in Ref. \cite{Sherson+10} was $0.07~U/k_{B}$.

\section{Outlook and Conclusions}

While at low values of $J/U$, $q$ clearly distinguishes between the BG and MI, the most obvious drawback of our work is the inability of $q$ to clearly identify a BG-SF transition. This is entirely due to the continuous chemical potential disorder we consider, such as is most straightforwardly added to current experiments by, for example, superimposing a speckle potential. If an alternative form of disorder was used instead that did not give rise to such significant inhomogeneities in the superfluid phase, such as bimodal mass disorder or hopping disorder, this order parameter would be capable of distinguishing both the MI-BG transition and the BG-SF transition.

The first measurement of an Edwards-Anderson order parameter in a strongly interacting quantum system would already be a landmark achievement, but may also allow the possibility of experimentally testing for replica symmetry breaking, a feat recently achieved in random laser systems \cite{Ghofraniha+15}. Replica symmetry breaking has been suggested to be crucial to the Bose glass \cite{Thomson+14,Thomson+15} but there is as yet no experimental confirmation of this. Future theoretical work going beyond the scope of the mean-field approach presented here could provide quantitative predictions of replica symmetry breaking for experiments to compare to.

While we have concentrated on the case of the Bose glass, the measurement of $q$ using site-resolved imaging is also applicable to other glassy phases. For example, ultracold atoms with cavity-mediated interactions offer the possibility to create interesting spin glass analogues \cite{Strack+11,Gopalakrishnan+11,Rotondo+15} where measurements of the Edwards-Anderson order parameter could prove to be extremely illuminating. Although our analysis was restricted to an ultracold gas of bosons, the measurements themselves are not: the protocol proposed here could equally well be applied to ultracold fermionic gases and it may be possible to generalize these techniques to scanning tunneling microscope systems.

Beyond the Edwards-Anderson order parameter, the quantum gas microscope is a promising tool for investigating other properties of glassy phases. For example, in a large enough system it would be possible to directly image the spacing of the rare superfluid regions within the BG, which could offer insights into the disputed percolation transition from BG to SF. In typical harmonic traps, the local chemical potential varies quickly across lattice sites, which restricts the BG to a small area, ruling out measurements of longer length-scale properties. The integration of quantum gas microscopes and spatial light modulators to create arbitrary potentials \cite{Zupancic+16, Choi+16} could be used to increase the size of the BG regions. In order to provide an almost flat potential with hard walls, the red-detuned optical lattice could be illuminated by a repulsive blue-detuned potential of comparable trap frequency to the lattice beams and an additional blue-detuned Laguerre-Gauss beam. Using holographic methods these blue-detuned potentials and the disorder could all be generated using a single spatial light modulator \cite{Bowman+15}. In such a trap with weakly-varying $\mu$, large samples of a single phase can be measured, allowing quantum gas microscopes to probe longer wavelength properties of the BG at a site-resolved level.

In summary, we have shown that current-generation quantum gas microscopes are capable of directly imaging the Bose glass phase. We have provided mean-field simulations of the Edwards-Anderson order parameter under realistic experimental conditions, paving the way for its first measurement in a strongly-correlated system, and suggested future directions for experiments. 

Supporting data for this work may be found in \cite{OpenData}.

\begin{acknowledgments}
We thank D Cassettari, A Daley, S Denny, J Keeling, P Kirton and A Trombettoni for insightful discussions and assistance. Computations were performed on the EPSRC CDT Computer Cluster and the University of St Andrews School of Physics \& Astronomy computer cluster. SJT acknowledges studentship funding from EPSRC under grant no. EP/G03673X/1. GDB acknowledges support from the Leverhulme Trust RPG-2013-074.
\end{acknowledgments}

\end{document}